\documentclass[11pt]{article}

\usepackage[margin=1in]{geometry}
\usepackage{graphicx} % support the \includegraphics command and options
\usepackage{amsmath} % for nice math commands and environments

\usepackage{amssymb} % for \mathbb, \mathfrak fonts
\usepackage{mathrsfs} % for \mathscr font
\usepackage{bm} % bold math symbols
\usepackage{units} %for nicefrac

\usepackage[font=footnotesize,labelfont=bf,labelsep=space]{caption} % advanced caption control; makes caption text smaller, makes figure labels bold, and adds space after label 
\usepackage{authblk} % advanced author control; make author affiliation footprints easy
\usepackage[squaren]{SIunits} % for nice units formatting e.g. \unit{50}{\kilo\gram}; squaren option prevents identical command error
\usepackage{hyperref} % produces hypertext links for equations, figures, sections, etc.
\usepackage{cite} % improved numerical citations
\usepackage{tensor} % enables commands \indices and \tensor for easily writing staggered indices

\newcommand{\scrA}{\mathcal{A}}
\newcommand{\scrB}{\mathcal{B}}

\mathchardef\mhyphen="2D

\newcommand{\vs}[1]{\rule{0pt}{#1ex}}

\usepackage[low-sup]{subdepth}

\begin{document}

\title{Classification of Electromagnetic and Gravitational Hopfions by Algebraic Type}

\author[1,2]{Amy Thompson}
\author[2,3]{Alexander Wickes}
\author[2,3]{Joe Swearngin}
\author[1,2]{Dirk Bouwmeester}

\affil[1]{\small Dept. of Physics, University of California, Santa Barbara, CA 93106}
\affil[2]{\small Huygens Laboratory, Leiden University, 2300 RA Leiden, The Netherlands}
\affil[3]{\small Dept. of Physics, University of California, Los Angeles, CA 90095}

\date{}

\maketitle

%------------------------------------------------
%------------------------------------------------
%------------------------------------------------
%------------------------------------------------

\begin{abstract}

We extend the definition of hopfions to include a class of spin-$h$ fields and use this to introduce the electromagnetic and gravitational hopfions of different algebraic types. The fields are constructed through the Penrose contour integral transform, thus the singularities of the generating functions are directly related to the geometry of the resulting physical fields. We discuss this relationship and how the topological structure of the fields is related to the Robinson congruence. Since the topology appears in the lines of force for both electromagnetism and gravity, the gravito-electromagnetic formalism is used to analyze the gravitational hopfions and describe the time evolution of their tendex and vortex lines. The correspondence between fields of different spin results in analogous configurations based on the same topological structure. The null and type N fields propagate at the speed of light, while the non-null and type D fields radiate energy outward from the center. Finally we discuss the type III gravitational hopfion, which has no direct electromagnetic analog, but find that it still exhibits some of the characteristic features common to the other hopfion fields.

\end{abstract}

%------------------------------------------------

%------------------------------------------------
\section{Introduction}
%------------------------------------------------

Since the earliest days of electromagnetism, when Gauss introduced his integral for calculating the linking number of two curves in 1833, the topology of physical fields has been of interest \cite{Nash1997book}. Around the same time, Faraday developed the concept of lines of force to give a direct, physical description of electromagnetic phenomena \cite{FaradayBook1870vol2}. By the turn of the 20th century ``tangles" or knots in solar currents were already being observed and could be explained by magnetic effects \cite{Bigelow1901,Mitchell1909}. Since then, the topological properties of lines of force, such as their linking, knotting, twisting, and kinking, have been studied in many physical systems \cite{Berger1984,Berger1999}.

Faraday had also hoped to find a similar intuitive description for gravitational lines of force. Although he and other scientists such as Maxwell gave much thought to finding this analogy connecting electromagnetism and gravity, they were never able to give an adequate explanation or mathematical formulation for this relationship \cite{FaradayBook2008vol5}. In modern times, the connection between massless linear relativistic fields of different spin has been understood in terms of the $SL(2,\mathbb{C})$ spinor field equations. In this language, the spin-$N$ equations - the Dirac, Maxwell, Rarita-Schwinger, and linear Einstein equations - take on a similar form \cite{Penrose1999central}. Their solutions can be expressed in terms of complex contour integrals allowing one to construct analogous fields based on topologically non-trivial configurations for any spin. 

The concept of field lines can be applied to gravity, as was once hoped by Faraday and Maxwell, by decomposing the curvature tensor into the forces experienced by a particular time-like observer allowing one to gain an intuitive understanding of the dynamics of curved space-time \cite{Nichols2011,Zhang2012,Nichols2012}. This gravito-electromagnetic analogy provides a direct way of analyzing linked and knotted fields, since the topology physically manifests in the lines of force for both electromagnetism and gravity. The topology of lines of force is also related to the dynamics of electromagnetic fields \cite{irvine2010linked, Arrayas2012exchange,Thompson2014plasma}. The spin-$N$ correspondence suggests that topology may influence the dynamics of gravitational fields and the gravito-electromagnetic analogy may provide an avenue for investigating such effects.

In this work, we will study fields with a topological structure based on the Hopf map, which is a surjective map sending great circles on $\mathbb{S}^3$ to points on $\mathbb{S}^2$, denoted by
\begin{equation}
\mathbb{S}^3 \overset{\mathbb{S}^1}{\longrightarrow}\mathbb{S}^2. \notag
\end{equation}
The great circles on $\mathbb{S}^3$ are called the fibers of the map, and when stereographically projected onto $\mathbb{R}^3$ correspond to the integral curves of physical fields. These circles lie on nested toroidal surfaces and each is linked with every other circle exactly once, creating the characteristic Hopf structure. The homotopy invariant of the map between two spheres is called the Hopf invariant, or Hopf index. This also corresponds to the topological invariant on $\mathbb{R}^3$, the linking number of the fibers. 

Traditionally, the term \textit{hopfion} has referred to soliton field configurations with Hopf index unity \cite{gladikowski1997static}. These kinds of hopfions have enjoyed a wide variety of application in many areas of physics \cite{Urbantke2003}. In particle physics, hopfions manifest as stable knot-like structures in field theoretic descriptions of hadrons \cite{Skyrme1962, Faddeev1997}. Hopfions have also been shown to represent localized topological solitons in several physical systems, including superfluid He phases \cite{Volovik1977}, spinor Bose-Einstein condensates \cite{Kawaguchi2008}, and MHD descriptions of plasma \cite{Kamchatnov1982}. It has also been shown that there exists a radiative solution, referred to as the EM hopfion, which is a solution to the vacuum Maxwell equations with the property that the field lines belonging to either the electric, magnetic, and Poynting vector fields have linking number one and lie on the surfaces of nested tori \cite{Ranada2002}.

An alternative construction for hopfion fields is based on the Robinson congruence, a null shear-free geodesic congruence that lies tangent to a Hopf fibration on all constant-time spatial foliations of Minkowski space. Thus a hopfion can be defined as a field configuration whose principal null directions (PNDs) all lie tangent to Robinson congruences. The twistor formalism provides a convenient method for obtaining these hopfions. Given a dual twistor\footnote{For our index conventions, we use lower case Latin letters for Lorentz indices, with $i$ and $j$ reserved for spatial indices, upper case Latin letters unprimed and primed for spinor and conjugate spinor indices respectively, and lower case Greek letters for twistor indices.} $A_\alpha = (\omega_A, \pi^{A'})$, there is an associated spinor field $\mathcal{A}_A = \omega_A - i x_{AA'} \pi^{A'}$ called the Robinson field whose flagpole is the Robinson congruence associated with the twistor \cite{Penrose1967}. This suggests that a hopfion may be represented entirely using twistors. 

Previously, we have used this formalism to construct the null EM hopfion solutions and find an analogous topologically non-trivial type N gravitational wave solution \cite{Swearngin2013}. Under time evolution, these radiative fields deform, but maintain their linked structure while propagating at the speed of light. In this work, we investigate how twistor methods can be used to find the spin-$h$ hopfion solutions of different spinor classifications. In particular, we will construct the non-null electromagnetic, type D and type III gravitational hopfion solutions, and then characterize the topology of their lines of force. The non-null and type D fields do not propagate, but radiate energy from the center of the configuration in all directions. The type III gravitational hopfion has no analogous EM counterpart, but can be written as a combination of the null and non-null hopfions. It also shares some of the distinctive traits of the other hopfions, for example they are localized, finite-energy fields with linked field line configurations.

%-----------------------------------------------
\section{Twistor Integral Methods}
%-----------------------------------------------

We consider the relationship between the singularities of the twistor generating functions and the geometry of the associated space-time fields. Since the fields are constructed through a contour integral transform, the poles of the twistor functions determine the spinor structure of the solutions in Minkowski space \cite{Penrose1977programme,Woodhouse1985methods}. By modifying the twistor function that generates the null EM hopfion, we find the generating function for a non-null EM solution with linked field lines. In a similar manner, we use the type N hopfion to construct gravitational hopfions of different Petrov classifications.

For our analysis, we will consider massless linear relativistic fields of helicity $h$, so their spin and helicity values are equivalent. Given a real-valued spin-$h$ field $\varphi_{A'_{1} \dots A'_{2h}}$ written in the self-dual $SL(2,\mathbb{C})$ spinor representation, the field can be decomposed into $2h$ single-index spinors called its principal spinors\cite{PenroseSpinors1}. This is described by the relation
\begin{equation}
\varphi_{{A'_1} \dots A'_{2h}} \propto \mathcal{A}_{(A'_1} \cdots \mathcal{D}_{A'_{2h})},
\end{equation}
with the one-index spinors representing the field's PNDs via their flagpole directions. Thus, for a hopfion arising from the definition given above, the PNDs of the spinors $\mathcal{A}, \dots, \mathcal{D}$ lie tangent to Robinson congruences. 

For example, a null spin-1 field can be written as
\begin{equation}
\varphi_{A'B'}(x) = f(x) \mathcal{A}_{A'}\mathcal{A}_{B'}
\end{equation}
where $\mathcal{A}_{A'}$ is an $SL(2,\mathbb{C})$ spinor which defines the doubly degenerate principal null direction of $F_{ab}$ and $f(x)$ is a scalar function that is determined by the electromagnetic field equation.
 
This suggests for any helicity $h$ there is an analogous spinor field
\begin{equation}
\varphi_{A'_1 \cdots A'_{2h}}(x) = f_{h}(x)\mathcal{A}_{A'_{1}}\cdots\mathcal{A}_{A'_{2h}}
\end{equation}
with $2h$-fold degenerate PNDs, but we will need to find the scaling function $f_h(x)$ that satisfies the appropriate spin-$h$ field equation.

The twistor formalism allows for the generalization of spin-1 fields to fields of any spin. The Penrose transform expresses solutions to the massless field equations as contour integrals over homogeneous twistor functions $F(Z)$
\begin{equation}
\label{PenroseTransform}
\varphi _{A_{1}'\cdots A_{2h}'}\left( x\right) 
=\frac{1}{2\pi i} \oint_\Gamma \pi _{A_{1}'} \cdots \pi_{A_{2h}'} F_{h}(Z) \pi_{B'}d\pi ^{B'}
\end{equation}
where $\Gamma$ is a contour on the Celestial sphere of $x$ and
\begin{equation}
F_{h}(\lambda Z) = \lambda^{-2h-2} F_{h}(Z)
\label{homogeneity}
\end{equation}
so that the degree of homogeneity is related to the helicity by $n_{hom}=-2h-2$ \cite{PenroseSpinors2}. In the Appendix, we give the explicit calculation of the non-null hopfion fields as an example to demonstrate how the pole structure of $F(Z)$ determines the corresponding field geometry through the contour integral in Eqn. \eqref{PenroseTransform}.

The twistor function which gives way to the EM hopfion through the Penrose transform is given by
\begin{equation}
\label{spin1twistorfunction}
F_{1}(Z) =\frac{1}{( A_{\alpha }Z^{\alpha }) ( B_{\beta }Z^{\beta })^{3}}.
\end{equation}
where the contour is taken around the pole defined by $A_{\alpha}$. To understand how the generating function relates to a particular space-time field configuration, consider the pole structure of Eqn. \eqref{spin1twistorfunction}. The term $(A \cdot Z)$ has a simple pole and the power of the $(B \cdot Z)$ term is chosen to give us homogeneity -4, and thus a spin-1 field. 

This approach leads directly to the spin-$h$ analogue. We keep the single pole for the $(A \cdot Z)$ term and the power of the $(B \cdot Z)$ term is determined by the relation between helicity and homogeneity $n_{hom}=-2h-2$ from Eqn. \eqref{homogeneity}. The generating function for the spin-$h$ hopfion is then
\begin{equation}
\label{Ntwistorfunction}
F_{h}(Z) =\frac{1}{( A_{\alpha }Z^{\alpha }) ( B_{\beta }Z^{\beta })^{2h+1}}.
\end{equation}
We find the associated spinor field from the Penrose transform of this twistor function is 
\begin{equation}
\varphi_{A'_1 \cdots A'_{2h}}(x) = \left( \frac{ 2 } {\Omega |x-y|^{2}} \right)^{2h+1}\mathcal{A}_{A'_{1}}\cdots\mathcal{A}_{A'_{2h}}
\label{null_phi}
\end{equation}
where $\Omega$ is a constant scalar, $y$ is a constant 4-vector determined by the specific values of ${A}_{\alpha}$ and ${B}_{\beta}$, and $\mathcal{A}_{A'}$  is the spinor field associated to the twistor $A_{\alpha}$ (see Appendix for definitions). As expected, we find the solution has a $2h$-fold degeneracy in its PNDs and the Penrose transform has given us the correct form of the scalar function that will satisfy the field equations.

%------------------------------------------------
\subsection{Petrov Variants} 
%------------------------------------------------

The twistor function in Eqn. \eqref{Ntwistorfunction} had a single pole which resulted in a spinor field with all its PNDs degenerate (null EM or type N gravity fields). Changing the pole structure yields solutions of different Petrov classes. Because the Penrose transform is a contour integral, when transforming functions with a pole of order greater than one Cauchy's integral formula involves the derivative of $F(Z)$. This derivative brings the other spinor field, in this case $\mathcal{B}_{B'}$, into the numerator thus breaking the degeneracy of the PNDs. For example, the twistor function
\begin{equation}
\label{eqn:twistor_function_nonnull_typeD}
F_{h}(Z) =\frac{1}{( A_{\alpha }Z^{\alpha })^{h+1} ( B_{\beta }Z^{\beta })^{h+1}}.
\end{equation}
results in non-null EM \{11\} or type D gravity \{22\} fields for $h=1,2$ given by
\begin{align}
\varphi_{A'_1 A'_2}(x) &= f_{1}(x) \mathcal{A}_{(A'_{1}} \mathcal{B}_{A'_2)} \\
\varphi_{A'_1 A'_2 A'_3 A'_4}(x) &= f_{2}(x)  \mathcal{A}_{(A'_{1}}\mathcal{A}_{A'_{2}} \mathcal{B}_{A'_{3}}\mathcal{B}_{A'_{4})}
\end{align}
where $\mathcal{A}_{A'}$ and  $\mathcal{B}_{A'}$  define the $h$-fold degenerate principal null directions. The details of the Penrose transform for the non-null spin-1 fields are given in the Appendix, and the other field types are found by a similar calculation. 

This gives all the classifications of the EM field strength tensor, which has two PNDs. For gravity we can also extend this to other classifications for which there is no EM analog, for example hopfion fields of Petrov type III are generated by
\begin{equation}
F_{2}(Z) =\frac{1}{( A_{\alpha }Z^{\alpha })^{2} ( B_{\beta }Z^{\beta })^{4}}.
\end{equation}
resulting in the field
\begin{equation}
\label{TypeIIItwistorfunction}
\varphi_{A'_1 \cdots A'_{4}}(x) = \left( \frac{ 2 } {\Omega |x-y|^{2}} \right)^{2h+1}\mathcal{A}_{(A'_{1}}\mathcal{A}_{A'_{2}}\mathcal{A}_{A'_{3}}\mathcal{B}_{A'_{4})}.
\end{equation}

%------------------------------------------------ 
%------------------------------------------------

In summary, the simplest hopfions correspond to homogeneous twistor functions of the form\footnote{These functions, first introduced by Penrose \cite{Penrose1972}, are a subset of the elementary states of twistor theory and play a fundamental role in solving problems in twistor space. For more discussion on the elementary states, see Refs. \cite{Penrose1987origins,Hodges1982diagrams,Eastwood1991density}.}
\begin{equation}
F(Z) = \frac{1}{(A \cdot Z)^{1+a} (B \cdot Z)^{1+b}},
\label{eqn:elementary_state}
\end{equation}
and the general solution is
\begin{equation}
\varphi_{A'_1 \dots A'_{2h}} = \left( \frac{ 2 } {\Omega |x-y|^{2}} \right)^{a+b+1} \scrA_{(A'_1} \cdots \scrA_{A'_b} \scrB_{A'_{b+1}} \cdots \scrB_{A'_{2h})}.
\label{eqn:elementary_hopfion}
\end{equation}
The spinor $\scrA_{A'}$ is the Robinson field of the twistor $A_\alpha$, and similarly $\scrB_{A'}$ is the corresponding Robinson field for the twistor $B_\beta$. Thus the hopfions of different algebraic type are characterized by two quantities $a$ and $b$, with $2h=a+b$. 

We now see that the null EM and null (type N) gravitational hopfions which we studied in \cite{Swearngin2013} are elementary hopfions with $a=0$. In fact all null hopfions take this form, as seen from the expression in Eqn. \eqref{eqn:elementary_hopfion} where, in the case $a=0$, the PNDs are all proportional to the flagpole of $\scrA_{A'}$ and thus completely degenerate.

%------------------------------------------------
\section{Electromagnetic Hopfions}
%------------------------------------------------

Before we consider the gravitational hopfions, we give a brief overview of the electromagnetic hopfions. The EM solutions are characterized by $h=1$, and thus have two distinct classifications: fields with two degenerate PNDs (null) and fields with two distinct PNDs (non-null). The null solution, originally due to Ra{\~n}ada \cite{Ranada2002}, can be derived using the pullback of the area element on $\mathbb{S}^{2}$ via the Hopf map to generate a field configuration with linked field lines. The alternative derivation using twistor generating functions presented here can be used to generalize the null solution to fields of different algebraic type based on the same topological structure. We use this method to construct the non-null EM hopfion and explore its properties.

%------------------------------------------------
\subsection{Null EM Hopfion}
%------------------------------------------------

\begin{figure}[t]
\begin{center}
\includegraphics[width=0.8\textwidth]{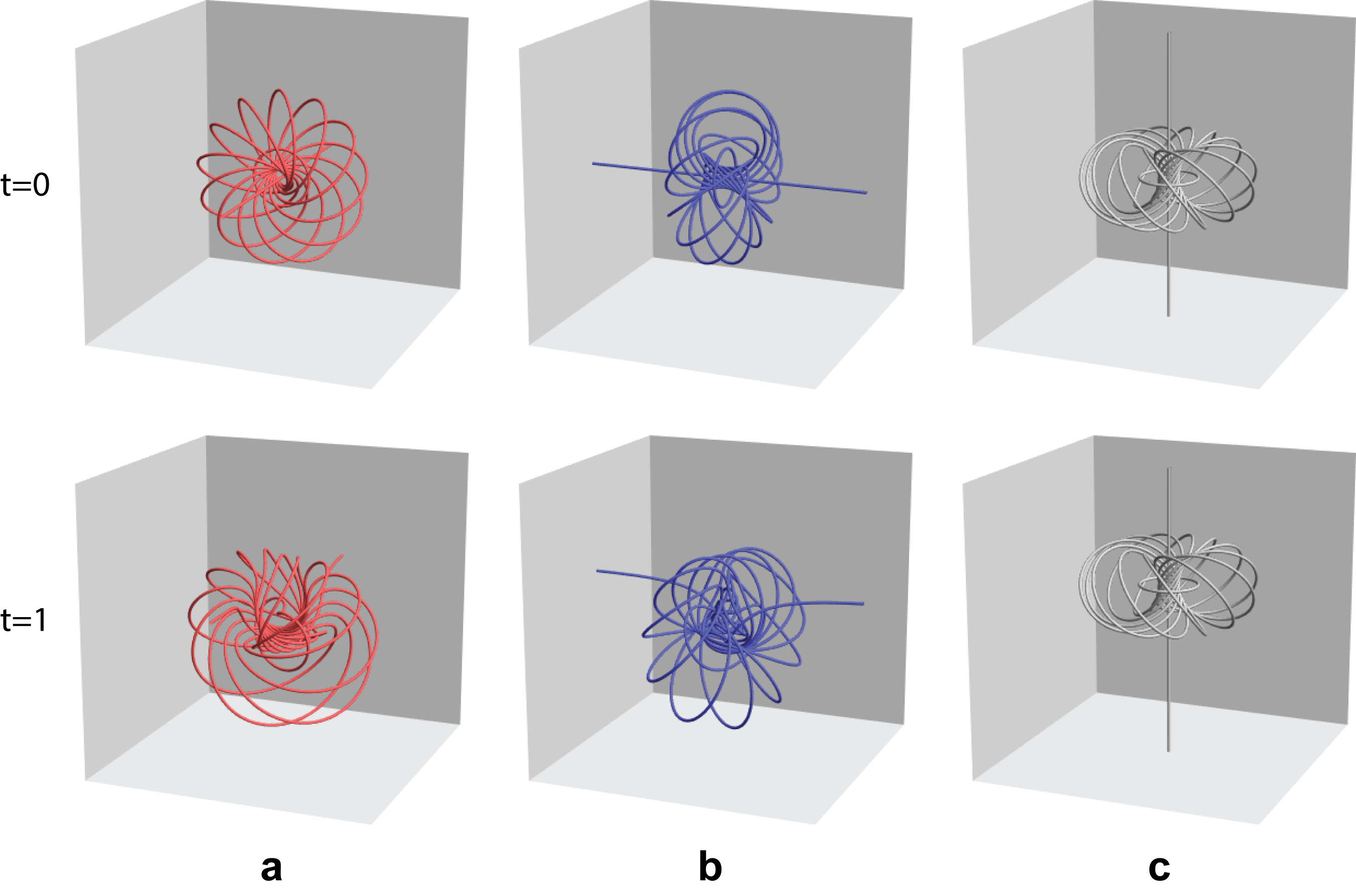}
\caption{Field line structure of the null EM hopfion: (\textbf{a}) the electric field, (\textbf{b}) the magnetic field, and (\textbf{c}) the Poynting vector. Row 1 shows the fields at $t=0$ are tangent to three orthogonal Hopf fibrations. The second row shows the $t=1$ configuration, where the electric and magnetic fields have deformed, but the Poynting vector is still everywhere tangent to a Hopf fibration and is propagating at the speed of light.}
\label{fig:EMhopfion}
\end{center}
\end{figure}

The null EM hopfion $\varphi_{A'B'} \sim \mathcal{A}_{A'} \mathcal{A}_{B'}$ is the simplest example of a hopfion, and exhibits a topologically non-trivial field line structure which is preserved as time evolves. It is a solution to the vacuum Maxwell equations such that any two field lines are closed and linked exactly once \cite{Besieris2009,Irvine2008}. When an EM hopfion is decomposed onto hyperplanes of constant time there always exists a hyperplane wherein the electric, magnetic, or Poynting vector fields are tangent to the fibers of three orthogonal Hopf fibrations \cite{van2013covariant}. The integral curves of the transport velocity, defined as the ratio of the Poynting vector to the electromagnetic energy density, do not deform but translate at the speed of light, thus preserving the topological structure of the field. As time evolves, the electric and magnetic fields deform, but the topology is conserved, so that the field lines still lie on the surfaces of nested tori and have linking number one. The field line structure is illustrated in Figure \ref{fig:EMhopfion}. 

To construct these fields from Eqn. \eqref{null_phi}, we must first make a choice of the dual twistors $A$ and $B$. The dual twistors
\begin{eqnarray*}
A_\alpha &=& (0,\sqrt{2},0,1) \\
B_\alpha &=& (\sqrt{2},0,1,0)
\end{eqnarray*}
correspond to Robinson congruences oriented in the $+z$ and $-z$ directions, respectively. The same choice of $A_\alpha$ and $B_\alpha$ is used in conjunction with Eqn. \eqref{eqn:elementary_hopfion} to construct the hopfion fields of different algebraic type.

The expressions for the electric and magnetic fields are complicated, but the solution takes a simple form when written as a Riemann-Silberstein (RS) vector 
\begin{align}
\label{eqn:null_em_RS}
\mathbf{F}_{\mathrm{null}} &= \mathbf{E} + i \mathbf{B} \\
	&= \frac{4}{\pi(-(t-i)^2+r^2)^3}
	\begin{pmatrix}
	(x-iz)^2 - (t-i+y)^2 \\ 
	2(x-iz)(t-i+y) \\ 
	i(x-iz)^2 + i(t-i+y)^2 
	\end{pmatrix}.
\end{align}
with $r^2=x^2+y^2+z^2$. We will analyze the gravitational solutions in terms of the properties of the EM hopfions, such as the energy density $u$ and Poynting vector $\mathbf{S}$. For the null case, these are given by
\begin{equation}
u_{\mathrm{null}} = \frac{1}{(d(x))^3}(1+x^2+y^2+(t-z)^2)^2,
\label{eqn:null_em_energy}
\end{equation}
\begin{equation}
\mathbf{S}_{\mathrm{null}} =  \frac{1}{(d(x))^3} (1 + x^2 + y^2 + (t-z)^2)
	\begin{pmatrix} 
	2(x(t-z)+y) \\
	2(y(t-z)-x) \\
	x^2 + y^2 - (t-z)^2 - 1
	\end{pmatrix}.
\label{eqn:null_em_poynting}
\end{equation}
The term $d(x)=(1 + 2 (t^2 + r^2) + (t^2 - r^2)^2)$ is a function related to the energy distribution of the field that appears in the expressions for all the hopfions. The overall scalar factor describes one of the quintessential features of these topological structures, namely that the energy density is concentrated at the center and falls off rapidly in the outward radial direction.

%------------------------------------------------
\subsection{Non-null EM Hopfion}
%------------------------------------------------

\begin{figure*}[t] % float placement: (h)ere, page (t)op, page (b)ottom, other (p)age
\centering
\includegraphics[width=0.92\textwidth]{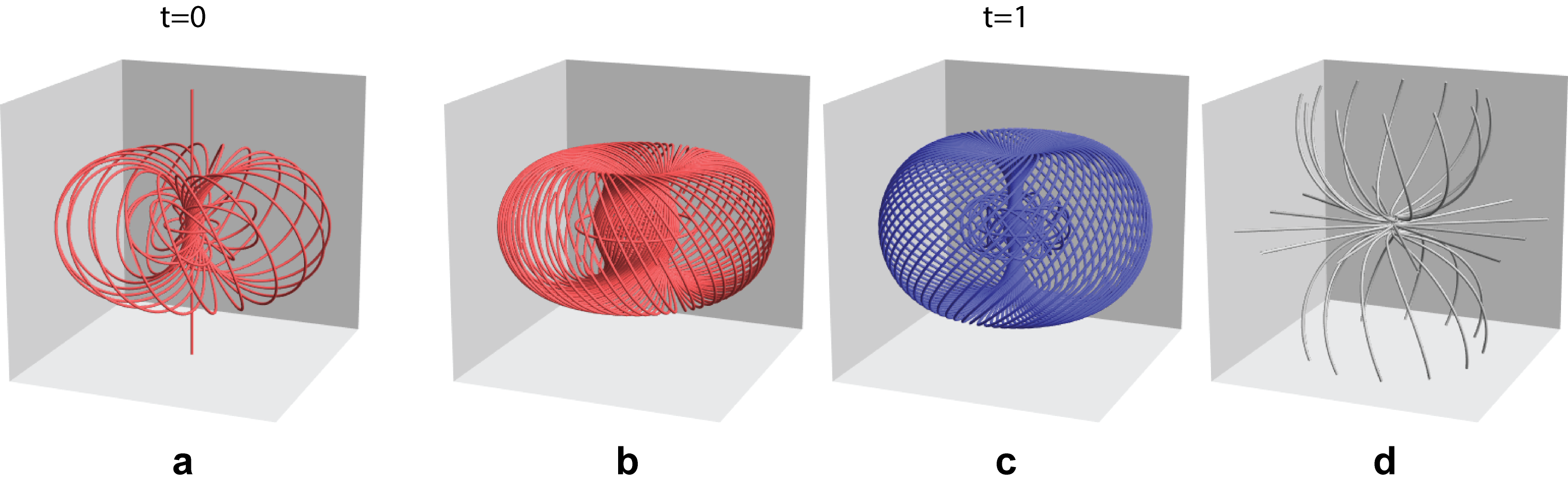}
\caption{Field line structure of the non-null EM hopfion. (\textbf{a}) The electric field lines at $t=0$ have the structure of the Hopf fibration. Not shown are $\mathbf{B}$ and $\mathbf{S}$, which are both identically zero at $t=0$. For $t=1$, (\textbf{b}) the electric field (\textbf{c}) the magnetic field, and (\textbf{d}) the Poynting vector.}
\label{fig:non-null_em}
\end{figure*}

The non-null EM hopfion $\varphi_{A'B'} \sim \scrA_{(A'} \scrB_{B')}$ can also be neatly expressed with a RS vector,
\begin{equation}
\mathbf{F}_{\mathrm{non \mhyphen null}} = \frac{2}{(-(t-i)^2+r^2)^3} 
	\begin{pmatrix}
	-2(xz - iy(t-i)) \\
	-2(yz + ix(t-i)) \\
	(t-i)^2 + x^2 + y^2 - z^2 
	\end{pmatrix}.
\label{eqn:non-null_em_RS}
\end{equation}
At $t=0$ the $\mathbf{E}$ field for this solution is everywhere tangent to a Hopf fibration, while $\mathbf{B}$ and hence $\mathbf{S}$ are identically zero since the RS vector is purely real. For $t \neq 0$, $\mathbf{B} \neq 0$ and the field line topologies for $\mathbf{E}$ and $\mathbf{B}$ are not preserved since $\mathbf{E} \cdot \mathbf{B} \neq 0$, however the fields are still finite energy. The field line structure of $\mathbf{S}$ shows that the solution does not propagate, but rather energy flows outward from the center of the configuration. These results are collected in Figure \ref{fig:non-null_em}. 

There are several known non-null EM configurations based on the Hopf map that are related to the fields in Eqn. \eqref{eqn:non-null_em_RS}. A solution found using the pullback method of Ra{\'n}ada is given in Ref. \cite{Arrayas2012exchange}. It is dual to fields described here, so that $\mathbf{F^{\prime}} \rightarrow -i \mathbf{F}_{\mathrm{non \mhyphen null}}$. Another configuration was constructed by Kiehn \cite{kiehn2002photon}, which is related to Eqn. \eqref{eqn:non-null_em_RS} by a transformation involving a complex shift in time and a global phase of $\pi/4$, given by 
\begin{equation}
\mathbf{F^{\prime}} \rightarrow \left. \tfrac{1}{\sqrt{2}}e^{\nicefrac{i \pi}{4}} \mathbf{F}_{\mathrm{non \mhyphen null}}\right| _{t \rightarrow t+i}.
\end{equation}

The non-null EM hopfion has the following expressions for the energy density and Poynting vector
\begin{equation}
u_{\mathrm{non \mhyphen null}} = \frac{2}{(d(x))^3} (t^4 + 2t^2(1+3r^2-4z^2) + (1+r^2)^2),
\label{eqn:non-null_em_energy}
\end{equation}
\begin{equation}
\mathbf{S}_{\mathrm{non \mhyphen null}} = \frac{4t}{(d(x))^3} 
	\begin{pmatrix} 
	x^3 + 2yz + x(1+t^2+y^2-z^2) \\ 
	y^3 - 2xz + y(1+t^2+x^2-z^2) \\
	2z(x^2+y^2) 
	\end{pmatrix}.
\label{eqn:non-null_em_poynting}
\end{equation}
We will see in the next section that these are related to the analogous gravitational hopfion fields.

%--------------------------------------
\section{Gravitational Hopfions}
%--------------------------------------

The gravitational hopfions are characterized by $h=2$, so there are a total of five distinct non-trivial gravitational hopfions, given by the Petrov types N, D, III, II, and I which classify the degeneracies of the PNDs. Here we review the type N, then present the type III and type D hopfions and analyze their structure using the GEM formalism.

%------------------------------------------------
\subsection{Review of GEM Tidal Tensors}
%------------------------------------------------

When analyzing the gravitational hopfions, it is useful to employ the gravito-electromagnetic (GEM) formalism, especially in cases where one can use this analogy to extract useful information about a solution from its electromagnetic counterpart.

For spin-1, the Penrose transform will generate a solution to the source-free field equation 
\begin{equation}
\nabla^{AA'}\varphi_{A'B'} = 0
\end{equation}
from which we construct the field strength tensor\footnote{We use the conventions of Penrose as in Ref. \cite{PenroseSpinors1}, so that $x^a \leftrightarrow x^{AA'} \equiv x^a \sigma^{AA'}_a = \frac{1}{\sqrt{2}} \begin{pmatrix} x^0 + x^3 & x^1 + ix^2 \\ x^1 - ix^2 & x^0 - x^3 \end{pmatrix}.$} by
\begin{align}
F_{A'B'AB} &= \varphi_{A'B'}\epsilon_{AB} + c.c. \notag \\
F_{ab} &= F_{A'B'AB}\sigma _{a}^{AA'}\sigma _{b}^{BB'}. 
\end{align}
For an observer at rest, we can decompose this into the standard electric and magnetic fields
\begin{align}
E_b &= F_{b0}, \\
B_b &= - \ast F_{b0}
\end{align}
The integral curves of $E_b$ and $B_b$ are the electromagnetic field lines.

For spin-2, the source-free field equation and Weyl curvature tensor are
\begin{align}
\nabla^{AA'}\varphi_{A'B'C'D'} &= 0, \notag\\ 
C_{A'B'C'D'ABCD} &= \varphi_{A'B'C'D'}\epsilon_{AB}\epsilon_{CD} + c.c. \notag \\
C_{abcd} &= C_{A'B'C'D'ABCD}\sigma _{a}^{AA'} \cdots \sigma _{d}^{DD'}.
\end{align}
In direct analogy with the decomposition of the electromagnetic field, the Weyl tensor $C_{abcd}$ can be decomposed into an even-parity ``electric" tensor $E_{ij}$ called the tidal field and an odd-parity ``magnetic" tensor $B_{ij}$ called the frame-drag field  \cite{Nichols2011}. For an observer at rest, these are
\begin{align}
E_{ij} &= C_{i0j0} \\
B_{ij} &= - \ast C_{i0j0}.
\end{align}
These tensors are symmetric and traceless, and are thus characterized entirely by their eigensystem configuration. The GE tensor has eigenvectors whose integral curves define the tendex lines and the eigenvalues $E_{\ell}$ are the magnitude of the tidal acceleration along these lines, where the relative acceleration over a small spatial separation $\ell$ is given by
\begin{align*}
\Delta a = -E_{\ell}\ell.
\end{align*}
Positive (negative) eigenvalues correspond to a compressive (stretching) force. The GM tensor has eigenvectors whose integral curves define the vortex lines and their eigenvalues $B_{\ell}$ are the magnitude of the gyroscope precession about the vortex lines 
\begin{align*}
\Delta\Omega = B_{\ell}\ell.
\end{align*}
Positive (negative) eigenvalues correspond to a clockwise (counter-clockwise) precession. The tendex and vortex lines are the gravitational lines of force for an observer at rest, which represent the analog of EM field lines.

In the GEM formalism, there are two local duality invariants analogous to the energy density and the Poynting vector in electromagnetism \cite{Maartens1998}. These are the super-energy density $U$ and the super-Poynting vector $\mathbf{P}$, given by
\begin{align}
U &= \frac{1}{2} (E_{ij} E^{ij} + B_{ij} B^{ij}) \\
P_i &= \varepsilon_{ijk}  E\indices{^j_l}B^{kl}.
\end{align}

%--------------------------------------
\subsection{Type N GEM Hopfion}
%--------------------------------------

The type N gravitational hopfion, previously studied in \cite{Swearngin2013}, has the form $\varphi_{A'B'C'D'} \sim \scrA_{A'} \scrA_{B'} \scrA_{C'} \scrA_{D'}$ and provides a good starting point for discussing the use of GEM in studying hopfions. For the type N hopfions, the Weyl decomposition has eigenvalues for both the GE and GM fields that take the form $\{ \lambda_-, \lambda_0, \lambda_+ \}$, with $\lambda_-(x) \leq \lambda_0(x) \leq \lambda_+(x)$ for all points $x$ in space-time. We will label the eigenvectors $\{\mathbf{e}_-, \mathbf{e}_0, \mathbf{e}_+\}$ and $\{\mathbf{b}_-, \mathbf{b}_0, \mathbf{b}_+\}$ corresponding to the eigenvalues for the tidal and frame-drag fields respectively. The type N fields are purely radiative, so the eigenvalues take the simple form $\{-\Lambda, 0, \Lambda\}$ where $\Lambda(x)$ is a function on space-time. 

\begin{figure}[t!] % float placement: (h)ere, page (t)op, page (b)ottom, other (p)age
\centering
\includegraphics[width=\textwidth]{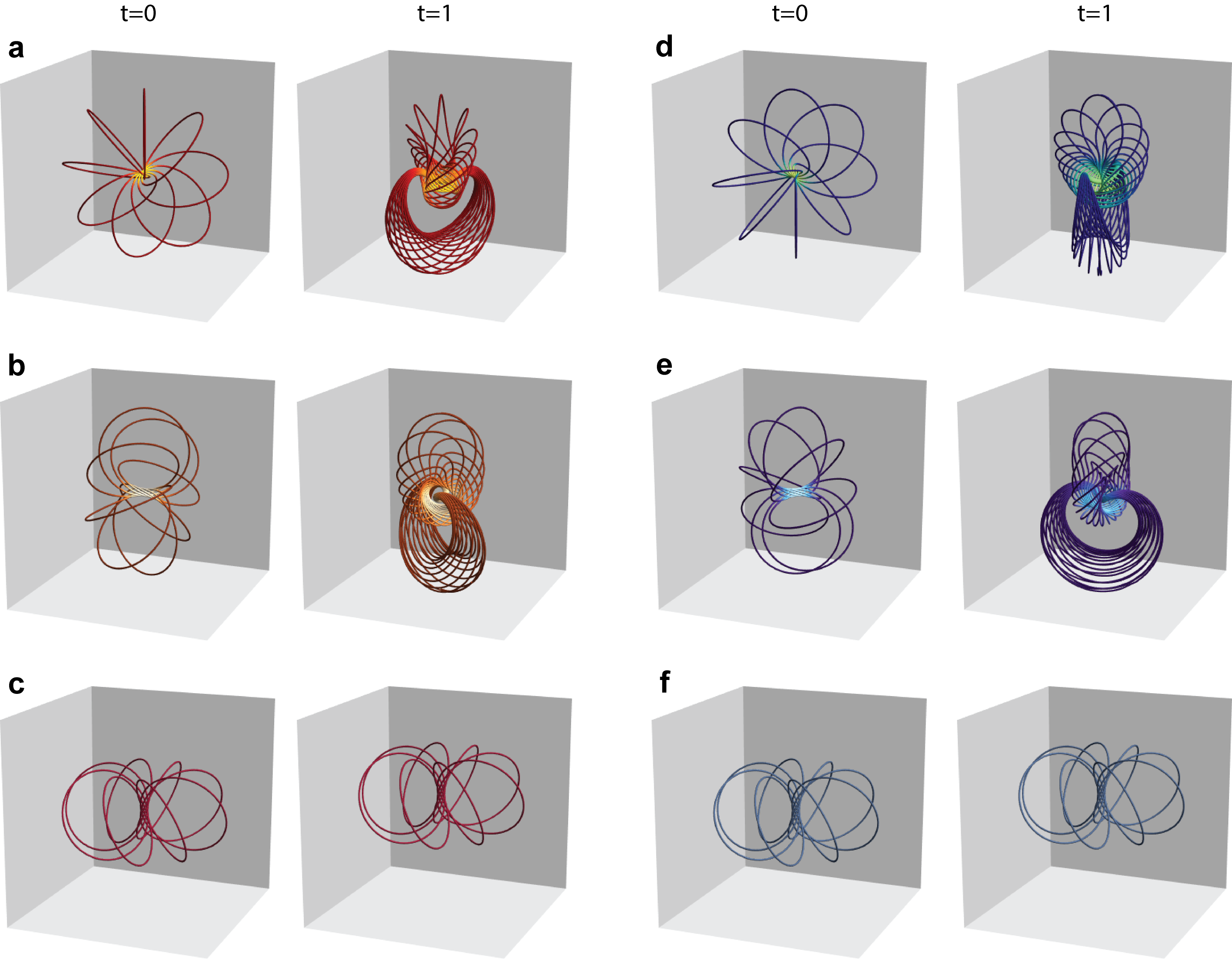}
\caption{The type N gravitational hopfion at $t=0$ and $t=1$: the tidal fields (\textbf{a}) $\mathbf{e}_-$, (\textbf{b}) $\mathbf{e}_+$, and (\textbf{c}) $\mathbf{e}_0$; and the frame-drag fields (\textbf{d}) $\mathbf{b}_-$, (\textbf{e}) $\mathbf{b}_+$, and (\textbf{f}) $\mathbf{b}_0$. The field lines are colored by the relative magnitude of their eigenvalues, with lighter colors indicating greater magnitude.}
\label{fig:typeN_grav}
\end{figure}

The eigenvectors $\mathbf{e}_0$ and $\mathbf{b}_0$ are both equivalent to the Poynting vector in Eqn. \eqref{eqn:null_em_poynting} for the null EM hopfion, up to an overall scalar. For the remaining eigenvector fields, we can construct RS vectors $\mathbf{f}_e = \mathbf{e}_- + i\mathbf{e}_+$ and $\mathbf{f}_b = \mathbf{b}_- + i\mathbf{b}_+$ which are related to the RS vector of the null EM hopfion from Eqn. \eqref{eqn:null_em_RS} by
\begin{align}
\mathbf{f}_e &= e^{i\pi/4} \mathbf{f}_b \\
&= e^{i \operatorname{Arg}\theta} \mathbf{F}_{\mathrm{null}}
\end{align}
where
\begin{equation}
\theta(x) = \sqrt{-(t-i)^2+r^2}.
\end{equation}
The tendex and vortex lines have been plotted numerically in Figure \ref{fig:typeN_grav}.

The super-energy and super-Poynting vector for this field are related to the duality invariants of the null EM hopfion by
\begin{equation}
U_{\vs{1.7} \mathrm{N}} = \frac{1}{2} d(x) u^2_{\mathrm{null}},
\end{equation}
\begin{equation}
\mathbf{P}_{\vs{1.7} \mathrm{N}} =  \frac{1}{2} d(x) u_{\mathrm{null}}\mathbf{S}_{\mathrm{null}}.
\end{equation}
Thus, we find the same Hopf structure that propagates at the speed of light. The surfaces of constant energy are concentric spheres, as in the spin-1 case, but the magnitude drops off more quickly in the radial direction.

%--------------------------------------
\subsection{Type D GEM Hopfion}
%--------------------------------------

\begin{figure*}[t!] % float placement: (h)ere, page (t)op, page (b)ottom, other (p)age
\centering
\includegraphics[width=0.8\textwidth]{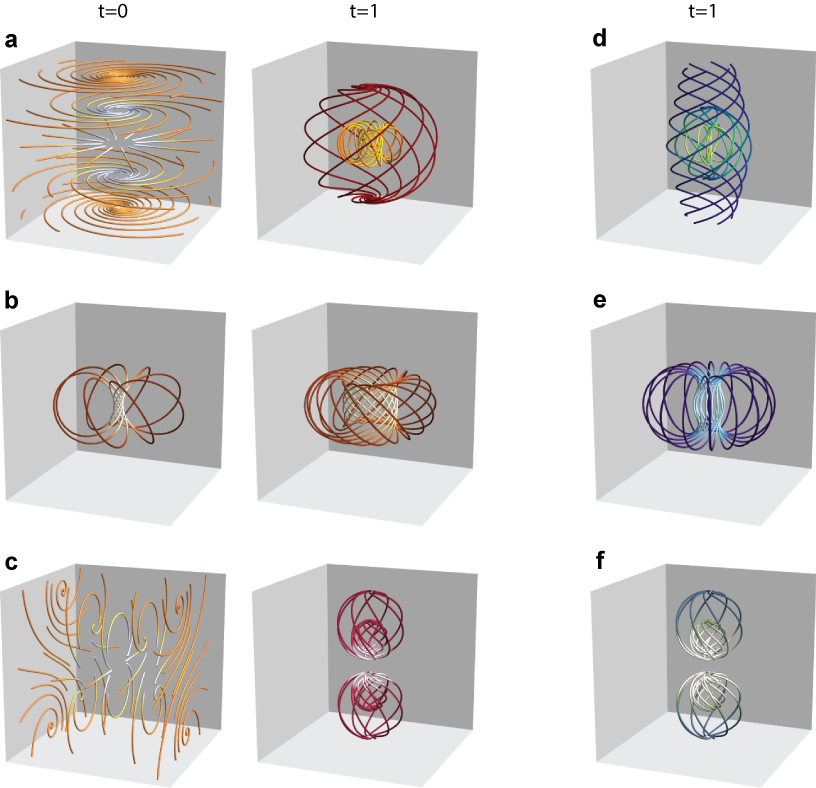}
\caption{The type D gravitational hopfion at $t=0$ and $t=1$: the tidal fields (\textbf{a}) $e_{-\Lambda+\lambda}$, (\textbf{b}) $e_{2\Lambda}$, and (\textbf{c}) $e_{-\Lambda-\lambda}$; and the frame-drag fields (\textbf{d}) $b_{-\Lambda+\lambda}$, (\textbf{e}) $b_{2\Lambda}$, and (\textbf{f}) $b_{-\Lambda-\lambda}$. The frame-drag fields at $t=0$ are omitted because they are all vanishing then. As before the field lines are colored by the relative magnitude of their eigenvalues, with lighter colors indicating greater magnitude. The fields at $t=0$ in (\textbf{a}) and (\textbf{c}) are presented with the same color scheme to convey the fact that they really represent a degenerate eigenspace together.}
\label{fig:typeD_grav}
\end{figure*}

The type D gravitational hopfion $\varphi_{A'B'C'D'} \sim \mathcal{A}_{(A'} \mathcal{A}_{B'} \mathcal{B}_{C'} \mathcal{B}_{D')}$ is the gravitational analog of the non-null EM hopfion, in that its PNDs are split evenly into two sets. For spin-2, the two sets consist of pairs of doubly degenerate PNDs. This hopfion has eigenvalue structure $\{2\Lambda, -\Lambda+\lambda, -\Lambda-\lambda\}$, where $\lambda=0$ at $t=0$ simplifying the eigenvalue structure to $\{2\Lambda, -\Lambda, -\Lambda\}$. Note that $\Lambda(x)$ used here represents a different function than the $\Lambda(x)$ used before to describe the type N hopfion; we use the symbol only to describe the overall structure of the eigenvalues. This eigenvalue configuration is interesting because at $t=0$ the eigenvalues $-\Lambda\pm\lambda$ coincide, so their eigenvectors collapse into a doubly degenerate eigenspace. Furthermore, at $t=0$, the GE field $\mathbf{e}_{2\Lambda}$ is exactly tangent to a Hopf fibration and the frame-drag field vanishes, hence the GM eigenvalues and eigenvectors vanish as well. The values of $\Lambda$ and $\lambda$ are rather complicated, so we will not present them here. The tendex and vortex lines have been plotted numerically in Figure \ref{fig:typeD_grav}.

%The values of $\Lambda$ and $\lambda$ are rather complicated, so we will not present them here. 

The expressions for the super-energy and super-Poynting vector of the type D hopfion are quite long, but they take a simpler form when written in terms of the duality invariants of the non-null EM hopfion
\begin{equation}
U_{\vs{1.7} \mathrm{D}} = \frac{1}{48} d(x) g_{\vs{1.7} \mathrm{D}}(x) u_{\mathrm{non \mhyphen null}}^2
\end{equation}
where
\begin{equation}
g_{\vs{1.7} \mathrm{D}}
(x) = \frac{(t^4 + 2 t^2 (1 + 5 r^2 - 6z^2) + (1 + r^2)^2)^2 - 
 48 t^4 (x^2 + y^2)^2}{(t^4 + 2 t^2 (1 + 3 r^2 - 4z^2) + (1 + r^2)^2)^2}
\end{equation}
and
\begin{equation}
\mathbf{P}_{\vs{1.7} \mathrm{D}} =  \frac{1}{32} d(x) u_{\mathrm{non \mhyphen null}} \mathbf{S}_{\mathrm{non \mhyphen null}}.
\end{equation}
Similarly to the non-null EM case, the super-Poynting vector indicates that the field configuration radiates energy outward from the center in all directions, but the overall structure does not propagate.

%--------------------------------------
\subsection{Type III GEM Hopfion}
%--------------------------------------

\begin{figure}[t!] % float placement: (h)ere, page (t)op, page (b)ottom, other (p)age
\centering
\includegraphics[width=\textwidth]{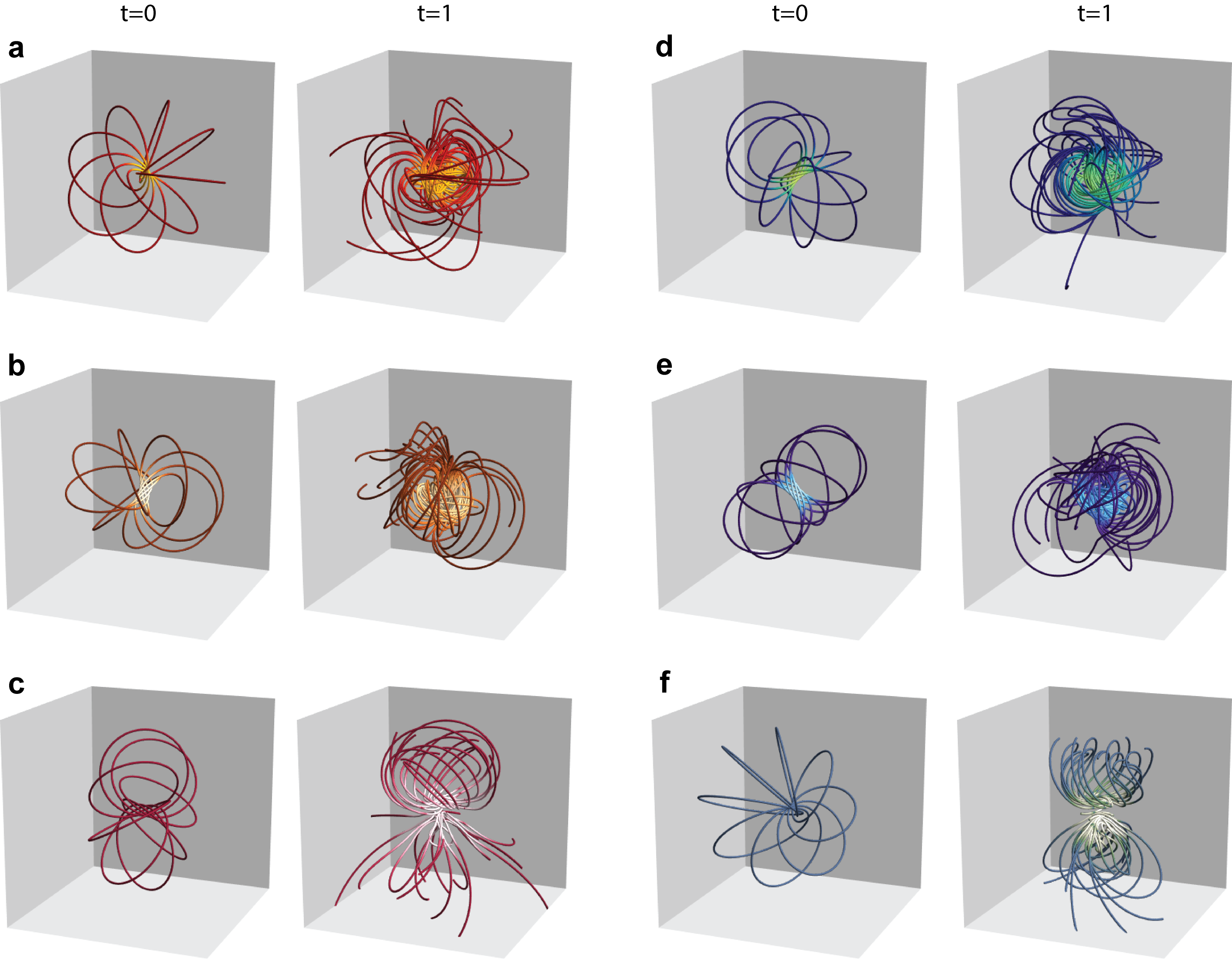}
\caption{The type III gravitational hopfion at $t=0$ and $t=1$: the tidal fields (\textbf{a}) $\mathbf{e}_-$, (\textbf{b}) $\mathbf{e}_+$, and (\textbf{c}) $\mathbf{e}_0$; and the frame-drag fields (\textbf{d}) $\mathbf{b}_-$, (\textbf{e}) $\mathbf{b}_+$, and (\textbf{f}) $\mathbf{b}_0$. The field lines are colored by the relative magnitude of their eigenvalues, with lighter colors indicating greater magnitude.}
\label{fig:typeIII_grav}
\end{figure}

The type III gravitational hopfion $\varphi_{A'B'C'D'} \sim \mathcal{A}_{(A'} \mathcal{A}_{B'} \mathcal{A}_{C'} \mathcal{B}_{D')}$ has one set of triply degenerate PNDs and one unique PND. This hopfion has eigenvalue structure $\{\lambda_-, \lambda_0, \lambda_+\} = \{-\Lambda, \lambda, \Lambda-\lambda\}$, where $\lambda=0$ at $t=0$. We again note that the functions $\Lambda(x)$ and $\lambda(x)$ used here represent different functions than those used to describe the type N and type D hopfions. At $t=0$, both the GE and GM fields are tangent to three orthogonal Hopf fibrations, but with different orientations than the type N configuration. For type III, the eigenvectors $\mathbf{e}_0$ and $\mathbf{b}_0$ are not aligned with the super-Poynting vector, but rather are orthogonal to it (and each other). The tendex and vortex lines have been plotted numerically in Figure \ref{fig:typeIII_grav}.

Comparing the type III hopfion to the EM hopfions, we see that the two sets of local duality invariants are similar. The super-energy is related by
\begin{equation}
U_{\vs{1.7} \mathrm{III}} = \frac{1}{16} d(x) g_{\vs{1.7} \mathrm{III}}(x) u_{\mathrm{null}
} u_{\mathrm{non \mhyphen null}}
\end{equation}
where
\begin{equation}
g_{\vs{1.7} \mathrm{III}}(x) = \frac{(t^4 + 2t^2(1+7r^2-8z^2) + (1+r^2)^2)}{(t^4 + 2t^2(1+3r^2-4z^2) + (1+r^2)^2)}.
\end{equation}
The super-Poynting vector can be written in terms of two vector terms that are exactly the same as the vector terms in the Poynting vectors of the EM fields, from Eqns. \eqref{eqn:null_em_poynting} and \eqref{eqn:non-null_em_poynting}, up to the overall scalar factors
\begin{equation}
\begin{split}
\mathbf{P}_{\vs{1.7} \mathrm{III}} &= \frac{1}{32} d(x) g_{\vs{1.7} \mathrm{III}}(x) u_{\mathrm{non \mhyphen null}} \mathbf{S}_{\mathrm{null}} \\
& \qquad \qquad \qquad + \frac{1}{16} d(x) u_{\mathrm{null}} \mathbf{S}_{\mathrm{non \mhyphen null}}.
\label{TypeIIIexplicit}
\end{split}
\end{equation}
Thus we see the field is comprised of two distinct structures. The first term in Eqn. \eqref{TypeIIIexplicit} corresponds to a component of the field that propagates at the speed of light, which is proportional to the Poynting vector of the null EM hopfion from Figure \ref{fig:EMhopfion}c. The second term is a component that radiates energy outward from the center, which is proportional to the Poynting vector of the non-null EM hopfion from Figure \ref{fig:non-null_em}d. The two terms combined create the configuration in \ref{fig:typeIII_grav}, so that
as time evolves the linked structure propagates in the $+\hat z$-direction, leaving open field lines radiating energy in the $-\hat z$-direction. This can be seen by comparing the visualizations of the EM Poynting vectors to the type III gravity fields in Figures \ref{fig:typeIII_grav}c and \ref{fig:typeIII_grav}f.

%------------------------------------------------
\subsection{Type II and Type I Fields}
%------------------------------------------------

Finally, we briefly mention the type II and type I fields. It is not possible to generate algebraically special fields of these types from a twistor function of the form in Eqn. \eqref{eqn:elementary_state}. Type II fields contain three distinct PNDs, therefore one must introduce a $(C \cdot Z)$ term into Eqn. \eqref{eqn:elementary_state}, where the twistor $C_\gamma$ has the associated spinor field $\mathcal{C}_{A'}$ which becomes one PND. However, when you apply Cauchy's integral theorem to the contour integral the derivative requires you use the product rule, thus the result includes multiple terms. The solution is then a linear combination of different type II fields. A similar situation arises for type I.

%------------------------------------------------
\section{Conclusion} 
%------------------------------------------------

The beauty of the Penrose transform lies in its complex contour integral nature, which allows for the application of Cauchy's theorem to bring out the spinor structure of solutions in $\mathbb{M}$. We used this method to modify the generating functions corresponding to the null EM and type N GEM hopfions and construct a class of spin-$h$ fields, including the non-null electromagnetic, type D and type III gravitational hopfions. The gravito-electromagnetic formalism was used to characterize the tendex and vortex structure of the gravitational hopfions and show that the linked configuration of the Hopf fibration appears in the lines of force. 

The fields based on the Hopf fibration studied here represent some of the most basic topological structures found in continuous, space-filling configurations. The methods we have presented could potentially be extended to the construction of classical electromagnetic and gravitational fields based on more intricate topological structures. For example, the radiative hopfions - the null electromagnetic and type N gravitational hopfions - have been shown to be the simplest case in a class of solutions with field line structures based on torus knots \cite{Irvine2013,Thompson2014GEMtorusknots}. After the torus knots, the twist knots are considered to be the next simplest class of knots  \cite{Rolfsen1976}, and have already been observed in other areas of physics such as polymer materials \cite{PikYin2002,Swetnam2012}, DNA organization \cite{Wasserman1991,Marenduzzo2009}, and quantum field theory \cite{Gukov2005,ElNaschie2008}. Identifying new field configurations and studying their properties could open new physical applications, and deepening our understanding of field line topology gives us insight into the structure and dynamics of physical systems.

%------------------------------------------------
\section*{Acknowledgments} 
%------------------------------------------------

The authors would like to thank J.W. Dalhuisen for helpful discussions. This work is supported by NWO VICI 680-47-604 and NSF Award PHY-1206118.

%-------------------------------------------------------------
\section*{Appendix: Penrose Transform for Non-null Hopfions}
%-------------------------------------------------------------

When the Penrose transform is written as an integral over a $\mathbb{CP}^1$ coordinate, the application of Cauchy's theorem to generating functions with poles of order greater than one results in derivative terms which break the degeneracy of the PNDs. Here we show the Penrose transform for the non-null spin-1 case, but the type D and type III spin-2 calculations follow in a similar manner. The calculation for the null and type N fields is given in Ref. \cite{Swearngin2013}.
 
We will calculate the Penrose transform
\begin{equation}
\phi (X)_{A'B'}=\frac{1}{2\pi i}\oint\limits_{\Gamma }\pi_{A'} \pi_{B'} F(Z)\pi _{B'} d\pi ^{B'} \notag
\end{equation}
with $f(Z)$ given by Eqn. \eqref{eqn:twistor_function_nonnull_typeD} with $h=1$
\begin{equation}
F(Z) = \frac{1}{(A_\alpha Z^\alpha)^{2} (B_\beta Z^\beta)^{2}}. \notag
\end{equation}
Let $A_\alpha$ and $B_\alpha $ be dual twistors associated to the spinor fields $\mathcal{A}^{A'}$ and $\mathcal{B}^{B'}$ according to
\begin{eqnarray*}
A_{\alpha }Z^{\alpha } &=&iA_{A}x^{AA'}\pi _{A'}+A^{A'}\pi _{A'} \notag\\
&\equiv &\mathcal{A}^{A'}\pi _{A'} \notag\\
B_{\beta }Z^{\beta } &=&iB_{B}x^{BB'}\pi _{B'}+B^{B'}\pi _{B'}\notag \\
&\equiv &\mathcal{B}^{A'}\pi _{B'},
\end{eqnarray*}
The measure can be written in the form
\begin{equation*}
\pi_{C'} d\pi^{C'} = (\pi_{0'})^2 d \zeta.
\end{equation*}
We also have the relations
\begin{align*}
\frac{1}{\pi_{0'}} \mathcal{A}^{A'} \pi_{A'} &= \mathcal{A}^{0'} + \mathcal{A}^{1'} \zeta, \\
	\frac{1}{\pi_{0'}} \mathcal{B}^{A'} \pi_{A'} &= \mathcal{B}^{0'} + \mathcal{B}^{1'} \zeta.
\end{align*}
Introducing the canonical spin bases $\{o_{A'}, \iota_{A'}\}$ into the primed spin space $S'$ we have that
\begin{align}
\pi_{A'} &= \pi_{0'} o_{A'} + \pi_{1'} \iota_{A'}  \notag \\
	&= \pi_{0'} (o_{A'} + (\frac{\pi_{1'}}{\pi_{0'}}) \iota_{A'}) \notag\\
	&= \pi_{0'} (o_{A'} + \zeta \iota_{A'}).
	\label{eqn:spin_bases}
\end{align}
Thus
\begin{align}
\varphi_{A' B'}(x) &= \frac{1}{2\pi i} \oint_\Gamma \frac { (o_{A'}+(\frac{\pi_{1'}}{\pi_{0'}}) \iota_{A'})(o_{B'}+(\frac{\pi_{1'}}{\pi_{0'}}) \iota_{B'})} {(\mathcal{A}^{0'}+\mathcal{A}^{1'}(\frac{\pi_{1'}}{\pi_{0'}}))^2(\mathcal{B}^{0'}+\mathcal{B}^{1'}(\frac{\pi_{1'}}{\pi_{0'}}))^2} d(\frac{\pi_{1'}}{\pi_{0'}}) \notag \\
&= \frac{1}{2\pi i (\mathcal{A}^{1'})^2 (\mathcal{B}^{1'})^2} \oint_\Gamma \frac{(o_{A'}+\zeta\iota_{A'})(o_{B'}+\zeta\iota_{B'})}{(\mu+\zeta)^2(\nu+\zeta)^2} d\zeta
\end{align}
where $\zeta = \pi_{1'} / \pi_{0'}$, $\mu = \mathcal{A}^{0'} / \mathcal{A}^{1'}$, and $\nu = \mathcal{B}^{0'} / \mathcal{B}^{1'}$ represent the projective coordinate and poles respectively.

After the variable substitutions, the integral is straightforward. Taking the contour $\Gamma$ to enclose the pole $-\mu$ and integrating each of the above terms in turn we arrive at 
\begin{eqnarray*}
\frac{1}{2 \pi i}\oint\limits_{\Gamma}o_{A'}o_{B'}\frac{d\zeta }{\left( \mu +\zeta \right)^2 \left( \nu +\zeta \right)^2} &=& o_{A'}o_{B'}\frac{2}{\left( \mu -\nu \right) ^{3}} \\
\frac{1}{2 \pi i}\oint\limits_{\Gamma }o_{A'}\imath _{B'}\frac{\zeta
d\zeta }{\left( \mu +\zeta \right)^2 \left( \nu +\zeta \right)^2}
&=& o_{A'}\imath _{B'}\frac{-(\mu +\nu)}{\left( \mu -\nu
\right)^3} \\
\frac{1}{2 \pi i}\oint\limits_{\Gamma }o_{B'}\imath _{A'}\frac{\zeta
d\zeta }{\left( \mu +\zeta \right)^2 \left( \nu +\zeta \right)^2}
&=& o_{B'}\imath _{A'}\frac{-(\mu +\nu)}{\left( \mu -\nu
\right)^3} \\
\frac{1}{2 \pi i}\oint\limits_{\Gamma }\imath _{B'}\imath _{A'}\frac{\zeta
^{2}d\zeta }{\left( \mu +\zeta \right)^2 \left( \nu +\zeta \right)^2}
&=& \imath _{B'}\imath _{A'}\frac{2\mu \nu} {\left(\mu -\nu \right)^3},
\end{eqnarray*}
which we reassemble to give the spinor field $\phi _{A'B'}$ as
\begin{eqnarray*}
\phi_{A'B'} &=&\frac{1}{\left( \mathcal{A}^{1'}\right)^2 \left(\mathcal{B}^{1'}\right)^2} \frac{1} {\left( \mu-\nu \right)^{3}} (2 o_{A'}o_{B'} - (\mu + \nu)(o_{A'}\imath_{B'} + o_{B'}\imath_{A'}) + 2 \mu \nu \imath_{B'} \imath_{A'}) \\
&=&\frac{( \mathcal{A}^{1'})( \mathcal{B}^{1'})}{(\varepsilon_{A'B'} \mathcal{A}^{A'} \mathcal{B}^{B'})^{3}} 
(2 o_{A'}o_{B'}- (\mu+\nu)(o_{A'}\imath_{B'} + o_{B'} \imath_{A'})+2 \mu \nu \imath_{B'}\imath _{A'}) \\
&=&\frac{( \mathcal{A}^{1'})( \mathcal{B}^{1'})} {(A_{A}B^{A}(x^a-y^a)(x_a-y_a))^{3}} 
(2 o_{A'}o_{B'}- (\mu+\nu)(o_{A'}\imath_{B'} + o_{B'} \imath_{A'})+2 \mu \nu \imath_{B'}\imath _{A'}) \\
&=&\frac{2}{(A_{A}B^{A}(|x-y|^2)^3}
(\mathcal{A}^{1'} \mathcal{B}^{1'}o_{A'}o_{B'} - \tfrac{1}{2}(\mathcal{A}^{0'} \mathcal{B}^{1'} + \mathcal{A}^{1'} \mathcal{B}^{0'})(o_{A'}\imath_{B'} + o_{B'} \imath_{A'}) +  \mathcal{A}^{0'} \mathcal{B}^{0'} \imath_{B'} \imath_{A'}) \\
&=&\frac{2}{(A_{A}B^{A}(|x-y|^2)^3}
(\mathcal{A}_{0'}\mathcal{B}_{0'} o_{A'}o_{B'}+\tfrac{1}{2}(\mathcal{A}_{1'} \mathcal{B}_{0'} + \mathcal{A}_{0'} \mathcal{B}_{1'})(o_{A'}\imath _{B'}+o_{B'}\imath _{A'})+ \mathcal{A}_{1'}\mathcal{B}_{1'} \imath _{B'}\imath _{A'}) \\
&=&\frac{2}{(\Omega|x-y|^2)^{3}} \mathcal{A}_{(A'} \mathcal{B}_{B')}
\end{eqnarray*}
where $\Omega = A_{A}B^{A}$ is a constant scalar and the point $y$ is given by
\begin{equation}
y^{AA'}=i\frac{A^{A'}B^{A}-B^{A'}A^{A}}{A_{B}B^{B}}.
\label{eqn:yAAprime}
\end{equation}


\begin{thebibliography}{10}

\bibitem{Nash1997book}
C.~Nash.
\newblock Topology and physics - a historical essay.
\newblock 1997, arXiv:hep-th/9709135.

\bibitem{FaradayBook1870vol2}
Bence Jones.
\newblock {\em The Life and Letters of Faraday, Volume 2}.
\newblock Longmans, Green, ans Co., London, 1870.

\bibitem{Bigelow1901}
Frank~H. Bigelow.
\newblock The magnetic theory of the solar corona.
\newblock {\em The American Journal of Science}, 11(64):253--262, 1901.

\bibitem{Mitchell1909}
Walter~M. Mitchell.
\newblock Recent solar observations at {Haverford}.
\newblock {\em The Astrophysical Journal}, 30(2):75--85, 1909.

\bibitem{Berger1984}
Mitchell Berger and George~B. Field.
\newblock The topological properties of magnetic helicity.
\newblock {\em Journal of Fluid Mechanics}, 147:133--148, 1984.

\bibitem{Berger1999}
Mitchell~A. Berger.
\newblock Introduction to magnetic helicity.
\newblock {\em Plasma Physics and Controlled Fusion}, 41(12B):B167, 1999.

\bibitem{FaradayBook2008vol5}
Michael Faraday and Frank A.~J.~L. James.
\newblock {\em The Correspondence of Michael Faraday, Volume 5: 1855-1860}.
\newblock Institution of Electrical Engineers, London, 2008.

\bibitem{Penrose1999central}
Roger Penrose.
\newblock The central programme of twistor theory.
\newblock {\em Chaos, Solitons \& Fractals}, 10(2):581--611, 1999.

\bibitem{Nichols2011}
David~A. Nichols, Robert Owen, Fan Zhang, Aaron Zimmerman, Jeandrew Brink,
  Yanbei Chen, Jeffrey~D. Kaplan, Geoffrey Lovelace, Keith~D. Matthews, Mark~A.
  Scheel, and Kip~S. Thorne.
\newblock Visualizing spacetime curvature via frame-drag vortexes and tidal
  tendexes: General theory and weak-gravity applications.
\newblock {\em Physical Review D}, 84(12):124014, 2011.

\bibitem{Zhang2012}
Fan Zhang, Aaron Zimmerman, David~A. Nichols, Yanbei Chen, Geoffrey Lovelace,
  Keith~D. Matthews, Robert Owen, and Kip~S. Thorne.
\newblock Visualizing spacetime curvature via frame-drag vortexes and tidal
  tendexes {II}. stationary black holes.
\newblock {\em Physical Review D}, 86(8):084049, 2012.

\bibitem{Nichols2012}
David~A. Nichols, Aaron Zimmerman, Yanbei Chen, Geoffrey Lovelace, Keith~D.
  Matthews, Robert Owen, Fan Zhang, and Kip~S. Thorne.
\newblock Visualizing spacetime curvature via frame-drag vortexes and tidal
  tendexes {III}. quasinormal pulsations of schwarzschild and kerr black holes.
\newblock {\em Physical Review D}, 86(10):104028, 2012.

\bibitem{irvine2010linked}
William T.~M. Irvine.
\newblock Linked and knotted beams of light, conservation of helicity and the
  flow of null electromagnetic fields.
\newblock {\em Journal of Physics A: Mathematical and Theoretical},
  43(38):385203, 2010.

\bibitem{Arrayas2012exchange}
Manuel Array{\'a}s and Jos{\'e}~L. Trueba.
\newblock Exchange of helicity in a knotted electromagnetic field.
\newblock {\em Annalen der Physik}, 524(2):71--75, 2012.

\bibitem{Thompson2014plasma}
Amy Thompson, Joe Swearngin, Alexander Wickes, and Dirk Bouwmeester.
\newblock Constructing a class of topological solitons in magnetohydrodynamics.
\newblock {\em Physical Review E}, 89(4):043104, 2014.

\bibitem{gladikowski1997static}
Jens Gladikowski and Meik Hellmund.
\newblock Static solitons with nonzero {Hopf} number.
\newblock {\em Physical Review D}, 56(8):5194, 1997.

\bibitem{Urbantke2003}
H.~K. Urbantke.
\newblock The {Hopf} fibration - seven times in physics.
\newblock {\em Journal of Geometry and Physics}, 46(2):125--150, 2003.

\bibitem{Skyrme1962}
T.~H.~R. Skyrme.
\newblock A unified field theory of mesons and baryons.
\newblock {\em Nuclear Physics}, 31(1):556--569, 1962.

\bibitem{Faddeev1997}
Ludvig Faddeev and Antti~J. Niemi.
\newblock Stable knot-like structures in classical field theory.
\newblock {\em Nature}, 387(6628):58--61, 1997.

\bibitem{Volovik1977}
G.~E. Volovik and V.~P. Mineev.
\newblock Particle-like solitons in superfluid {He} phases.
\newblock {\em Sov. Phys. JETP}, 46(2):401--404, 1977.

\bibitem{Kawaguchi2008}
Yuki Kawaguchi, Muneto Nitta, and Masahito Ueda.
\newblock Knots in a spinor {Bose-Einstein} condensate.
\newblock {\em Physical Review Letters}, 100(18):180403, 2008.

\bibitem{Kamchatnov1982}
A.~M. Kamchatnov.
\newblock Topological solitons in magnetohydrodynamics.
\newblock {\em Sov. Phys. JETP}, 55(1):69--73, 1982.

\bibitem{Ranada2002}
Antonio~F. Ra\~{n}ada and Jos{\'e}~L. Trueba.
\newblock Topological electromagnetism with hidden nonlinearity.
\newblock {\em Modern Nonlinear Optics, Part {III}}, 119:197--253, 2002.

\bibitem{Penrose1967}
Roger Penrose.
\newblock Twistor algebra.
\newblock {\em Journal of Mathematical Physics}, 8(2):345--366, 1967.

\bibitem{Swearngin2013}
Joe Swearngin, Amy Thompson, Alexander Wickes, Jan~Willem Dalhuisen, and Dirk
  Bouwmeester.
\newblock Gravitational hopfions.
\newblock 2014, arXiv:gr-qc/1302.1431.

\bibitem{Penrose1977programme}
Roger Penrose.
\newblock The twistor programme.
\newblock {\em Reports on Mathematical Physics}, 12(1):65--76, 1977.

\bibitem{Woodhouse1985methods}
N.~M.~J. Woodhouse.
\newblock Real methods in twistor theory.
\newblock {\em Classical and Quantum Gravity}, 2(3):257--291, 1985.

\bibitem{PenroseSpinors1}
Roger Penrose and Wolfgang Rindler.
\newblock {\em Spinors and Space-Time: Volume 1, Two-Spinor Calculus and
  Relativistic Fields}.
\newblock Cambridge Monographs on Mathematical Physics. Cambridge University
  Press, Cambridge, 1987.

\bibitem{PenroseSpinors2}
Roger Penrose and Wolfgang Rindler.
\newblock {\em Spinors and Space-Time: Volume 2, Spinor and Twistor Methods in
  Space-Time Geometry}.
\newblock Cambridge monographs on mathematical physics. Cambridge University
  Press, Cambridge, 1988.

\bibitem{Penrose1972}
Roger Penrose and M.~A.~H. MacCallum.
\newblock Twistor theory: An approach to the quantisation of fields and
  space-time.
\newblock {\em Physics Reports}, 6(4):241--316, 1972.

\bibitem{Penrose1987origins}
Roger Penrose.
\newblock On the origins of twistor theory.
\newblock In W.~Rindler and A.~Trautman, editors, {\em Gravitation and
  geometry, a Volume in Honour of I. Robinson}. Bibliopolis, Naples, 1987.

\bibitem{Hodges1982diagrams}
Andrew Hodges.
\newblock Twistor diagrams.
\newblock {\em Physica A: Statistical Mechanics and its Applications},
  114(1):157--175, 1982.

\bibitem{Eastwood1991density}
M.~G. Eastwood and A.~M. Pilato.
\newblock On the density of twistor elementary states.
\newblock {\em Pacific Journal of Mathematics}, 151(2):201--215, 1991.

\bibitem{Besieris2009}
Ioannis~M. Besieris and Amr~M. Shaarawi.
\newblock {Hopf-Ran{\~a}da} linked and knotted light beam solution viewed as a
  null electromagnetic field.
\newblock {\em Optics Letters}, 34(24):3887--3889, 2009.

\bibitem{Irvine2008}
William T.~M. Irvine and Dirk Bouwmeester.
\newblock Linked and knotted beams of light.
\newblock {\em Nature Physics}, 4(9):716--720, 2008.

\bibitem{van2013covariant}
S.~J. Van~Enk.
\newblock The covariant description of electric and magnetic field lines of
  null fields: application to {Hopf--Ra{\~n}ada} solutions.
\newblock {\em Journal of Physics A: Mathematical and Theoretical},
  46(17):175204, 2013.

\bibitem{kiehn2002photon}
Robert~M. Kiehn.
\newblock The photon spin and other topological features of classical
  electromagnetism.
\newblock In {\em Gravitation and Cosmology: From the Hubble Radius to the
  Planck Scale}, pages 197--206. Springer, 2002.

\bibitem{Maartens1998}
Roy Maartens and Bruce~A. Bassett.
\newblock Gravito-electromagnetism.
\newblock {\em Classical and Quantum Gravity}, 15(3):705, 1998.

\bibitem{Irvine2013}
Hridesh Kedia, Iwo Bialynicki-Birula, Daniel Peralta-Salas, and William M.~T.
  Irvine.
\newblock Tying knots in light fields.
\newblock {\em Physical Review Letters}, 111(15):150404, 2013.

\bibitem{Thompson2014GEMtorusknots}
Amy Thompson, Joe Swearngin, and Dirk Bouwmeester.
\newblock Linked and knotted gravitational radiation.
\newblock {\em Journal of Physics A: Mathematical and Theoretical},
  47(35):355205, 2014.

\bibitem{Rolfsen1976}
Dale Rolfsen.
\newblock {\em Knots and Links}.
\newblock AMS Chelsea Publishing, Providence, RI, 1976.

\bibitem{PikYin2002}
Pik-Yin Lai.
\newblock Dynamics of polymer knots at equilibrium.
\newblock {\em Physical Review E}, 66(2):021805, 2002.

\bibitem{Swetnam2012}
Adam~D. Swetnam, Charles Brett, and Michael~P. Allen.
\newblock Phase diagrams of knotted and unknotted ring polymers.
\newblock {\em Physical Review E}, 85(3):031804, 2012.

\bibitem{Wasserman1991}
Steven~A. Wasserman and Nicholas~R. Cozzarelli.
\newblock Supercoiled {DNA}-directed knotting by {T4} topoisomerase.
\newblock {\em The Journal of Biological Chemistry}, 266(30):20567--73, 1991.

\bibitem{Marenduzzo2009}
Davide Marenduzzo, Enzo Orlandini, Andrzej Stasiak, De~Witt Sumners, Luca
  Tubiana, and Cristian Micheletti.
\newblock {DNAâ€“-DNA} interactions in bacteriophage capsids are responsible
  for the observed {DNA} knotting.
\newblock {\em Proceedings of the National Academy of Sciences},
  106(52):22269--74, 2009.

\bibitem{Gukov2005}
Sergei Gukov.
\newblock Three-dimensional quantum gravity, {Chern-Simons} theory, and the
  {A}-polynomial.
\newblock {\em Communications in Mathematical Physics}, 255(3):577--627, 2005.

\bibitem{ElNaschie2008}
M.~S.~El Naschie.
\newblock Fuzzy multi-instanton knots in the fabric of spaceâ€“time and
  {Dirac}'s vacuum fluctuation.
\newblock {\em Chaos, Solitons \& Fractals}, 38(5):1260--1268, 2008.

\end{thebibliography}
\end{document}